\documentclass[prd,aps,preprintnumbers, nofootinbib,superscriptaddress, notitlepage]{revtex4}
\usepackage{epsfig}
\usepackage{bm}
\usepackage{amssymb}
\usepackage{amsmath}
\usepackage{color}
\usepackage{subfigure}
\usepackage[colorlinks,
            linkcolor=blue,
            anchorcolor=black,
            citecolor=blue
            ]{hyperref}

\allowdisplaybreaks[4]

\newcommand{\lperp}{\ensuremath{l_\perp}}

\newcommand{\kperp}{\ensuremath{k_\perp}}

\newcommand{\dd}{\ensuremath{\mathrm{d}}}

\begin{document}

\title{A parton shower generator based on the GLR equation}

\author{Yu Shi} %\email{yu.shi@sdu.edu.cn} 
\affiliation{Key Laboratory of Particle Physics and Particle Irradiation (MOE), Institute of Frontier and Interdisciplinary Science, Shandong University, Qingdao, Shandong 266237, China}

\author{Shu-Yi Wei}% \email{shuyi@sdu.edu.cn} 
\affiliation{Key Laboratory of Particle Physics and Particle Irradiation (MOE), Institute of Frontier and Interdisciplinary Science, Shandong University, Qingdao, Shandong 266237, China}

\author{Jian Zhou}%\email{jzhou@sdu.edu.cn}
\affiliation{Key Laboratory of Particle Physics and Particle Irradiation (MOE), Institute of Frontier and Interdisciplinary Science, Shandong University, Qingdao, Shandong 266237, China}

\begin{abstract}
We develop a novel Monte Carlo parton branching algorithm based on the Gribov-Levin-Ryskin (GLR) equation. The formulations of both forward evolution and backward evolution for the GLR equation are presented. The results from the Monte Carlo implementation of the GLR equation are in full agreement with its numerical solutions. Our work thus paves the way for developing an event generator that embodies the saturation effect.
\end{abstract}
\maketitle

\section{Introduction}

The Monte Carlo event generator is an indispensable tool for describing the exclusive hadronic final states of high energy scattering processes involving multi-particle production. One of the essential elements of modern general-purpose event generators is the simulation of a succession of emissions from the incoming and outgoing partons or the colored dipoles~\cite{Catani:1996vz,Jung:2000hk,Jung:2010si,Buckley:2011ms,Hoche:2014rga,Li:2016yez,Cabouat:2017rzi,Campbell:2021svd,Aschenauer:2022aeb,vanBeekveld:2022zhl,BermudezMartinez:2022bpj, Byer:2022bqf}. The majority of implementations of the parton branching process are built on the soft and collinear approximation, which allows us to effectively resum the Dokshitzer-Gribov-Levin-Altarelli-Parisi (DGLAP) like logarithm to all orders by an iteration procedure. 

Several alternative approaches, like SMALLX~\cite{Marchesini:1990zy, Marchesini:1992jw}, CASCADE~\cite{Jung:2000hk,Jung:2010si} and High Energy Jets exclusive partonic Monte Carlo (HEJ)~\cite{Andersen:2009nu, Andersen:2009he,Andersen:2011hs}, have been developed to include the contribution of semi-hard emissions which give rise to the Balitsky-Fadin-Kuraev-Lipatov (BFKL)~\cite{Kuraev:1977fs,Balitsky:1978ic} type large logarithm in the small $x$ region. These modern algorithms are quite sophisticated implementations of all-order perturbative QCD. In the formulation of CASCADE, the semi-hard gluon emissions are generated according to the Catani-Ciafaloni-Fiorani-Marchesini (CCFM) evolution equation~\cite{Ciafaloni:1987ur,Catani:1989sg,Catani:1989yc,Marchesini:1994wr}, while the resummation in HEJ is achieved by directly computing the hard matrix elements in the limit of large invariant mass between all particles to all orders. 

In the small $x$ region of a large nucleus target where gluon density is extremely high, the non-linear process of gluon-gluon recombination that limits the density growth becomes as important as the gluon branching process. A new semi-hard scale, the so-called saturation scale $Q_s$ dynamically emerges at small $x$. At this scale, the density of gluons is expected to saturate as gluon splitting and recombination reach a balance. To account for this non-linear effect, one has to go beyond the BFKL type evolution and simulate both the gluon splitting and gluon recombination process simultaneously. The objective of this study is to develop a Monte Carlo branching algorithm incorporating the saturation effect, which can describe the fully exclusive hadronic final states in $eA$ collisions at the Electron-Ion Collider (EIC)~\cite{Accardi:2012qut,AbdulKhalek:2021gbh}.

There are several nonlinear extensions of the BFKL equation, among which the most general one is the Jalilian-Marian–Iancu–McLerran–Weigert–Leonidov–Kovner (JIMWLK)~\cite{ JalilianMarian:1997jx, JalilianMarian:1997gr, Iancu:2000hn, Ferreiro:2001qy} evolution equation. Owing to the JIMWLK's complexity, it is a formidable task to implement it in a parton shower generator. On the other hand, what has been most widely used in phenomenology studies is the JIMWLK's mean field approximation version: the Balitsky-Kovchegov (BK)~\cite{Balitsky:1995ub,Kovchegov:1999yj} equation. However, the BK equation does not form a good basis for a parton shower generator either, since the Fourier transform of the dipole amplitude entering the BK equation in the momentum space lacks a clear probability interpretation. We will elucidate this point in more details in the next section. It turns out that the GLR equation~\cite{Gribov:1983ivg}, which contains a nonlinear damping term resulting from the double BFKL ladder mergers, is well suited to computer implementation. As compared to the BK equation, all higher order ($3\rightarrow 1$, $4\rightarrow 1$, etc.) multiple-ladder recombination is neglected in the GLR equation. However, the GLR equation remains a good approximation at intermediate energy where the $2\rightarrow 1$ gluon fusion dominates, and thus should be sufficient for simulating the events in $eA$ collisions at the EIC energy.

The rest of the paper is organized as follows. In Sec.~II, we first derive the folded version of the GLR equation from its standard form. The folded GLR equation is the starting point for realizing Monte Carlo implementation. In Sec.~III, we present the formulation of the forward evolution for the GLR equation starting with the discussion about the dilute limit case, i.e., the BFKL equation. It is shown that the $k_\perp$ distribution obtained in the forward evolution approach is in full agreement with the numerical solutions of the BFKL equation and the GLR equation respectively. In Sec.~IV, we discuss the algorithm for backward evolution that is necessary for any practical phenomenology applications. We confirm that the parton cascade generated in the backward evolution approach is identical to that obtained from the forward evolution. The paper is summarized in Sec.~V. 

\section{The folded and the unfolded GLR equation}

Before discussing the Monte Carlo implementation of the GLR equation, let us first explain why it is difficult to build a BK-based parton shower generator. The BK equation describes the rapidity evolution of the two-point correlation function which is also referred to as the dipole scattering amplitude. The BK equation in momentum space is most conveniently expressed in terms of a Fourier transform of the dipole amplitude multiplying with the factor $1 /r_\perp^2$,
\begin{equation}
\mathcal N( \eta,\kperp)=\int\frac{\dd^{2}r_{\perp}}{2\pi}
\frac{e^{-i\kperp\cdot r_{\perp} }}{r_\perp ^2}
\left[ 1-\frac{1}{N_c} \langle U^\dag(0) U(r_\perp) \rangle 
\right ] ,
\end{equation}
where $U(r_\perp)=\mathcal P \exp \left[ ig\int dz^- A^+(z^-,r_\perp) \right ]$ is a lightlike Wilson line in the fundamental representation. The rapidity $\eta$ is defined as $\eta=\ln \left(x_0/x\right) $ with $x_0=0.01$. In terms of $\mathcal N$, the BK equation reads~\cite{Kovchegov:1999ua,Marquet:2005zf},
\begin{eqnarray}
\frac{\partial \mathcal N(\eta,\kperp)}{\partial \eta }&=&
 \frac{\bar \alpha_s}{\pi} \left[ \int\frac{\dd^{2}\lperp}{\lperp^{2}} 
\mathcal N( \eta, l_\perp+ k_\perp) -
\int_0^{k_\perp} \frac{\dd^{2}\lperp}{\lperp^{2}} \mathcal N( \eta,\kperp) \right]- \bar \alpha_{s} \mathcal N^{2}( \eta,\kperp),\label{eq:1-1}
\end{eqnarray}
with $\bar \alpha_s= \alpha_s N_c/\pi$. The first two linear terms in Eq.~(\ref{eq:1-1}), which coincide with those in the BFKL kernel, correspond to contributions from the real and virtual gluon emissions respectively. Here, we present the virtual correction in a form~\cite{Kwiecinski:1996td,Kutak:2011fu} that is different from the conventional expression. The equivalence between the two forms is shown in Appendix A. The last term is the nonlinear term arising from the resummation of fan diagrams. One can solve the BK equation and obtain the distribution $\mathcal N$ at arbitrary rapidity $\eta$ using the algorithm described below. However, there exists no clear probability interpretation for the distribution $\mathcal N$. The gluon branching constructed with $\mathcal N$ from Monte Carlo simulation thus does not correspond to a real parton cascade. Furthermore, from the point of view of a sensible description of exclusive quantities, it is not only the evolved gluon distribution that matters. In deriving the BK equation, all the radiated gluons have been integrated out. In this way, all multiple-point correlation functions, which show up in the intermediate steps of the derivation, eventually collapse into the two-point function. On the other hand, one has to explicitly keep the four momenta of all radiated gluons in a parton shower generator. If the emitted gluons were left unintegrated, the multiple-point correlation functions~\cite{Blaizot:2004wv, Dominguez:2008aa,Dominguez:2011wm,Dominguez:2012ad, Shi:2017gcq, Zhang:2019yhk} beside the dipole one will enter the evolution equation. One should use the JIMWLK equation to simulate the parton branching process instead. Therefore, we conclude that the BK equation does not form a good basis for a parton shower generator. 

Now let us turn to discuss the GLR equation. The GLR evolution equation introduced in Ref.~\cite{Gribov:1983ivg} was one of the first few attempts~\cite{Gribov:1983ivg,Mueller:1985wy} to tackle the BFKL unitarity problem by including a quadratic damping term resulting from the $2\rightarrow 1$ gluon fusion process. It is directly expressed in terms of the unintegrated gluon distribution~\cite{Gribov:1983ivg,Bartels:2007dm},
\begin{eqnarray}
\frac{\partial G(\eta,k_{\perp})}{\partial \eta } \!\!&=&\!\! \frac{\bar \alpha_s}{\pi} \left [ \int \frac{\dd^2 l_{\perp}}{l_\perp^2} G(\eta,k_\perp\!+\!l_\perp)
- \int_0^{k_\perp} \frac{\dd^2 l_{\perp}}{l_\perp^2} G(\eta,k_{\perp}) \right ] - g_{\rm TPV} \frac{\alpha_s^2}{S_\perp (8\pi)^2}G^2(\eta,k_{\perp}), \label{GLR1}
\end{eqnarray}
where $S_\perp$ denotes the transverse area of the target. $g_{\rm TPV}$ is an effective coupling constant resulting from the local approximation of the triple pomeron vertex~\cite{Bartels:1994jj,Bartels:2007dm}. By requiring the GLR equation and the BK equation to coincide with each other in the dilute limit, we fix this effective coupling constant to be $g_{\rm TPV}=8(2\pi)^4$. Different values of $g_{\rm TPV}$ could be derived depending on how one treats the triple pomeron vertex. $G(\eta,k_{\perp})$ is the transverse momentum dependent (TMD) gluon distribution describing the gluon number density for a given $k_\perp$ and $\eta$. There are two different types of gluon TMDs widely used in phenomenological studies~\cite{Kharzeev:2003wz,Dominguez:2011wm}: the dipole gluon distribution and the Weizsacker-Williams (WW) gluon distribution. Their small-$x$ evolutions are governed by the BK equation and the Dominguez-Mueller-Munier-Xiao (DMMX) equation~\cite{Dominguez:2011gc}, respectively. In the moderate small $x$ region where the triple-pomeron-vertice contribution dominates over other higher-order effects, the evolution of both gluon TMDs is expected to be described by the GLR equation approximately. 
 
To facilitate the following algebraic manipulations, we cast Eq.~(\ref{GLR1}) into the following form with the replacement $N(\eta,\kperp)=\frac{2\alpha_s \pi^3}{ N_c S_\perp} G(\eta,k_{\perp})$,
\begin{eqnarray}
\frac{\partial N(\eta,k_{\perp})}{\partial \eta } \!\!&=&\!\! \frac{ \bar \alpha_s }{\pi} \left [ \int \frac{\dd^2 l_{\perp}}{l_\perp^2} N(\eta,k_\perp\!+\!l_\perp)
 - \int_0^{k_\perp} \frac{\dd^2 l_{\perp}}{l_\perp^2} N(\eta,k_{\perp}) \right ] -\bar \alpha_s N^2(\eta,k_{\perp}).
\end{eqnarray}
By making the identification $\mathcal N(\eta,\kperp)= N(\eta,\kperp)$~\cite{Kovchegov:1999ua}, the above equation is the same as the BK equation in Eq.~(\ref{eq:1-1}). However, we emphasize that this is nothing but merely a coincidence. Though the identification $\mathcal N(\eta,\kperp)= N(\eta,\kperp)$ can be shown to be valid in the dilute region, there is no exact relation between them in the region where multiple re-scattering and quantum evolution are important. 

Following the common procedure of implementing the DGLAP-based Monte Carlo algorithm, we have to construct a function describing the probability of evolving from $\eta_i$ to $\eta_{i+1}$ without resolvable branching and gluon fusion. To do so, we first separate the real correction into two terms as following,
\begin{eqnarray}
\int \frac{\dd^2 l_{\perp}}{l_\perp^2} N(\eta,k_\perp\!+\!l_\perp) \! =\!\! \int_\mu \!\frac{\dd^2 l_{\perp}}{l_\perp^2} N(\eta,k_\perp\!+\!l_\perp) +\!\int_0^\mu \! \frac{\dd^2 l_{\perp}}{l_\perp^2} N(\eta,k_\perp\!+\!l_\perp) \!\approx\!\! \int_\mu \! \frac{\dd^2 l_{\perp}}{l_\perp^2} N(\eta,k_\perp\!+\!l_\perp) +\!\int_0^\mu \! \frac{\dd^2 l_{\perp}}{l_\perp^2} N(\eta,k_\perp),
\end{eqnarray}
where the infrared cutoff $\mu$ is a matter of choice of what we classify as a resolvable emission. Branchings in the regime of $l_\perp<\mu $ are classified as unresolvable since they involve the emission of an undetectable soft gluon. The emissions beyond this region are classified as resolvable branchings. The next step is to combine the contribution from the unresolvable real emission with that from the virtual diagrams. We obtain
\begin{eqnarray}
\frac{\partial N( \eta,\kperp)}{\partial \eta }
&=&\frac{\bar \alpha_s}{\pi} \int_\mu \frac{\dd^{2}\lperp}{\lperp^{2}} N(\eta, l_\perp+ k_\perp ) 
- \bar \alpha_s \ln \frac{k_\perp ^2}{\mu ^2} N(\eta ,\kperp) - \bar \alpha_s N^{2}(\eta ,\kperp). \label{unfolded}
\end{eqnarray}
By introducing an auxiliary function $\Phi(\eta,k_\perp)$, $N(x,k_{\perp})$ can be expressed as
\begin{eqnarray}
N(\eta,k_{\perp}) = \Phi(\eta,k_\perp)\Delta (\eta , k_\perp),
\end{eqnarray}
where 
\begin{eqnarray}
\Delta (\eta , k_\perp) = \exp \left \{-\bar \alpha_s \int^\eta_{\eta_0} d\eta' \left [ \ln\frac{k_\perp^2}{\mu^2} +N(\eta',k_{\perp}) \right ] \right \}.
\end{eqnarray}
According to Eq.~(\ref{unfolded}), the function $\Phi(\eta,k_\perp)$ satisfies the following equation 
\begin{eqnarray}
\Delta (\eta , k_\perp) \frac{\partial\Phi(\eta,k_\perp) }{\partial \eta} 
\!&=&\! \frac{ \bar \alpha_s }{\pi} \int_\mu \frac{\dd^2 l_{\perp}}{l_\perp^2} N(\eta,k_\perp\!+\!l_\perp).
\end{eqnarray}
The above equation can be re-expressed in terms of $N(\eta,k_\perp) $ as
\begin{equation}
\frac{\partial}{\partial \eta } \frac{ N(\eta,\kperp)}{\Delta (\eta, k_\perp) } = \frac{ \bar \alpha_s }{\pi} \int_\mu \frac{\dd ^2 l_\perp }{l_\perp ^2} \frac{N(\eta,\lperp+\kperp)}{\Delta (\eta, k_\perp)}, \label{folded}
\end{equation}
which is referred to as the folded GLR equation, while Eq.~(\ref{unfolded}) or Eq.~(\ref{GLR1}) is the unfolded version. In the folded GLR equation, the unresolvable real emissions and the virtual correction have been manifestly resummed to all orders. $\Delta (\eta , k_\perp)$ represents the probability of evolving from $\eta_0$ to $\eta$ without a resolvable branching or gluon fusion. It reduces to the non-Sudakov form factor~\cite{Kwiecinski:1996td,Kutak:2011fu} in the small $x$ limit with the saturation term being neglected. Eq.~(\ref{folded}) can be integrated over to give an integral equation for $N(\eta ,\kperp)$. It reads
\begin{equation}
N(\eta ,\kperp)= N(\eta_0 ,\kperp) \Delta (\eta, k_\perp)
+ \frac{\bar \alpha_s}{\pi} \int^{\eta}_{\eta_0 }\dd \eta^\prime \frac{\Delta (\eta, k_\perp)}{\Delta (\eta^\prime, k_\perp)} \int _\mu \frac{\dd ^2 l_\perp }{l_\perp ^2} N(\eta^\prime,\lperp+\kperp), \label{eq:inbk}
\end{equation}
where $N(\eta_0 ,\kperp)$ is the gluon distribution at the initial rapidity. 

Small $x$ evolution equations resum the leading logarithmic contributions in terms of $\ln (1/x)$. However, from both theoretical and phenomenological points of view, the necessity of resuming the next-to-leading logarithmic contributions has long been recognized. There are several sources that give rise to the sub-leading logarithmic contributions, such as the running coupling effect~\cite{Kovchegov:2006vj, Balitsky:2006wa, Gardi:2006rp, Albacete:2007yr,Balitsky:2007feb,Berger:2010sh}, kinematic constraint~\cite{Kwiecinski:1996td, Kwiecinski:1997ee,Beuf:2014uia, Deak:2019wms, Liu:2022xsc}, the collinear improvement of the BK equation~\cite{Avsar:2011ds, Iancu:2015vea, Iancu:2015joa, Lappi:2016fmu,Ducloue:2019ezk, Ducloue:2019jmy}, and the Sudakov suppressed BK kernel~\cite{Zheng:2019zul}. Though these corrections are formally sub-leading power contributions, they often have a significant impact on the observables of interest at small $x$. We only discuss the Monte Carlo implementation of the running coupling effect in this work and leave the implementation of other effects for future works. It is quite straightforward to include the running coupling effect for the case of parent dipole prescription, which we will adopt in this study. It is not trivial to introduce kinematic constraint in the GLR equation. Following the arguments made in Refs.~\cite{Kwiecinski:1996td,Deak:2019wms}, the transverse momentum square of the radiated gluon $l_\perp^2$ must be smaller than $\frac{1-z}{z}k_\perp^2$ where $k_\perp$ and $z$ are transverse momentum and longitudinal momentum fraction carried by the daughter gluon respectively. The inclusion of such kinematic constraint leads to a modified GLR equation, which is given by
\begin{eqnarray}
\frac{\partial N(\eta,\kperp)}{\partial \eta }&=&
 \frac{\bar \alpha_s}{\pi} \int\frac{\dd^{2}\lperp}{\lperp^{2}} 
 N\left ( \eta +\ln \frac{ k_\perp^2}{ k_\perp^2+ l_\perp^2}, l_\perp+k_\perp \right ) -
\frac{\bar \alpha_s}{\pi} \int_0^{k_\perp} \frac{\dd^2 l_{\perp}}{l_\perp^2} N(\eta,k_{\perp}) - \bar \alpha_{s} N^{2}( \eta,\kperp).
\end{eqnarray}
Converting the above equation to the folded form of the GLR equation, we obtain
\begin{equation}
\frac{\partial}{\partial \eta } \frac{N(x,\kperp)}{\Delta (\eta, k_\perp) } = \frac{ \bar \alpha_s }{\pi} \int _\mu \frac{\dd ^2 l_\perp }{l_\perp ^2} \frac{
 N\left ( \eta +\ln \frac{ k_\perp^2}{ k_\perp^2+ l_\perp^2} ,\lperp+\kperp \right ) }{\Delta (\eta , k_\perp)}.
 \label{eq:bk_kc}
\end{equation}
The implementation of the kinematic constraint in the parton branching algorithm turns out to be quite non-trivial. We will address this in a separate publication. In this work, we focus on developing the Monte Carlo algorithms based on the folded evolution equations presented in Eq.~(\ref{folded}) and Eq.~(\ref{eq:inbk}).

\section{Forward evolution}

To demonstrate the formulation of forward evolution for the GLR equation, we start with the simplest case, i.e., the Monte Carlo implementation of the fixed coupling BFKL evolution. All essential elements of the algorithm will be discussed in this simplest example. The first step is to sample the $k_\perp$ distribution at the initial rapidity $\eta_0=0$ using the MV model~\cite{McLerran:1993ni, McLerran:1993ka} result as the input. Since we aim at building an event generator for $eA$ collisions, it is natural to use the WW type gluon distribution as the initial condition, which is given by,
\begin{align}
N (\eta=0, k_\perp) = \int \frac{d^2 r_\perp}{2\pi} e^{-i k_\perp \cdot r_\perp} \frac{1}{r_\perp^2} \left(1- \exp \bigl[-\frac{1}{4} Q_{s0}^2 r_\perp^2 \ln(e+\frac{1}{\Lambda r_\perp}) \bigr] \right),
\end{align}
with $Q_{s0}^2 = 1$ GeV$^2$ and $\Lambda=0.24$ GeV. To efficiently generate an event with this initial condition, we use a veto algorithm (see Appendix B for more details). Since the evolution variable is the rapidity, the basic problem one has to solve is that given ($\eta_i$, $k_{\perp, i}$) after some steps of the evolution, or given the initial condition, generating the values ($\eta_{i+1}$, $k_{\perp, i+1}$) in the next step. The Monte Carlo implementation is laid out in the following:

I): The first quantity to be generated by the algorithm is the value of $\eta_{i+1}$. One can read the probability of evolving from $\eta_i$ to $\eta_{i+1}$ without a resolvable branching from the folded BFKL equation. It is given by $\Delta (\eta_i, \eta_0; k_{\perp,i})/\Delta (\eta_{i+1}, \eta_0; k_{\perp,i})$. Thus $\eta_{i+1}$ can be generated with the correct probability distribution by solving the following equation, 
\begin{equation}
\mathcal R_1 = \exp\left[- \bar \alpha_s \int_{\eta_i} ^{\eta_{i+1}} \dd \eta ^\prime \ln \frac{ k_{\perp,i}^2}{\mu ^2} \right], \label{14}
\end{equation}
with the saturation effect being neglected in the BFKL case. Throughout this paper, we use ${\cal R}_i$ to represent a random number distributed uniformly in the interval [0,1].

II): The second step is to generate the value of radiated gluon's transverse momentum with a probability distribution proportional to $\bar \alpha_s \int \frac{\dd^2 l_\perp}{l_\perp^2}$, which is the real part of the BFKL kernel. It can be achieved by solving the following equation for $|l_\perp|$,
\begin{equation}
\mathcal{R}_2 \int^{P_\perp}_\mu \frac{\dd^2 l_{\perp}'}{l_\perp'^2} = \int_\mu ^{|l_\perp|} \frac{\dd^2 l_{\perp}'}{l_\perp'^2},
\end{equation}
where $P_\perp$ is the UV cut-off for the emitted gluon's transverse momentum. 

III): The azimuthal angle of $l_\perp$ is sampled according to,
\begin{align}
2\pi \mathcal{R}_3 =\phi_l.
\end{align}

IV): The minus component of the radiated gluon's momentum is obtained using the on-shell condition. The four momentum of the next exchanged gluon is reconstructed according to $k_{i+1}=k_{i}-l$.

V): The generated cascade needs to be re-weighted. The re-weighting factor associated with this branching is given by,
\begin{equation}
{\cal W} ( k_{\perp,i}) =\frac{ \ln(P_\perp^2/\mu^2) }{ \ln(k_{\perp,i}^2/\mu^2)},
\end{equation}
such that the number of exchanged gluons increases after each splitting. This is because of the mismatch between the phase spaces of the integrations for real and virtual corrections. For a given rapidity interval $\Delta \eta$, the number of gluons which vanish due to the virtual correction is proportional to $\Delta \eta \bar \alpha_s \int_{\mu}^{k_{\perp,i}} \frac{\dd l_\perp^2}{l_\perp^2} \exp \left[-\bar \alpha_s (\eta_i-\eta_{i+1})\ln\frac{k_{\perp,i}^2}{\mu^2} \right]$, while the number of gluons produced via the real correction is proportional to $\Delta \eta \bar \alpha_s \int_{\mu}^{P_\perp} \frac{\dd^2 l_\perp }{l_\perp^2} \exp \left[-\bar \alpha_s (\eta_i-\eta_{i+1})\ln\frac{k_{\perp,i}^2}{\mu^2}\right]$ in the same rapidity interval. The re-weighting function is given by the ratio of these two contributions.

\begin{figure}[htb]
\centering
\includegraphics[width=0.4\textwidth]{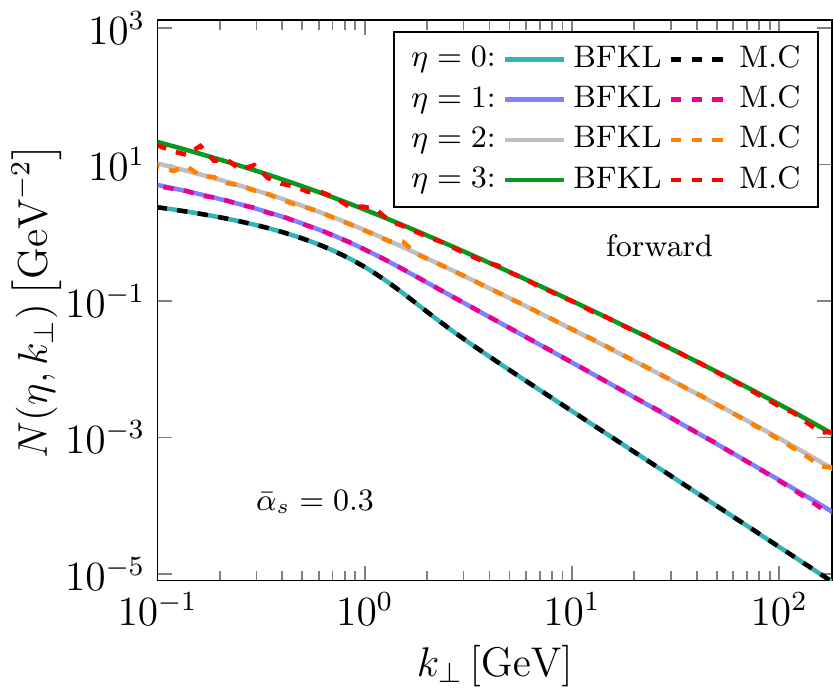}
\includegraphics[width=0.4\textwidth]{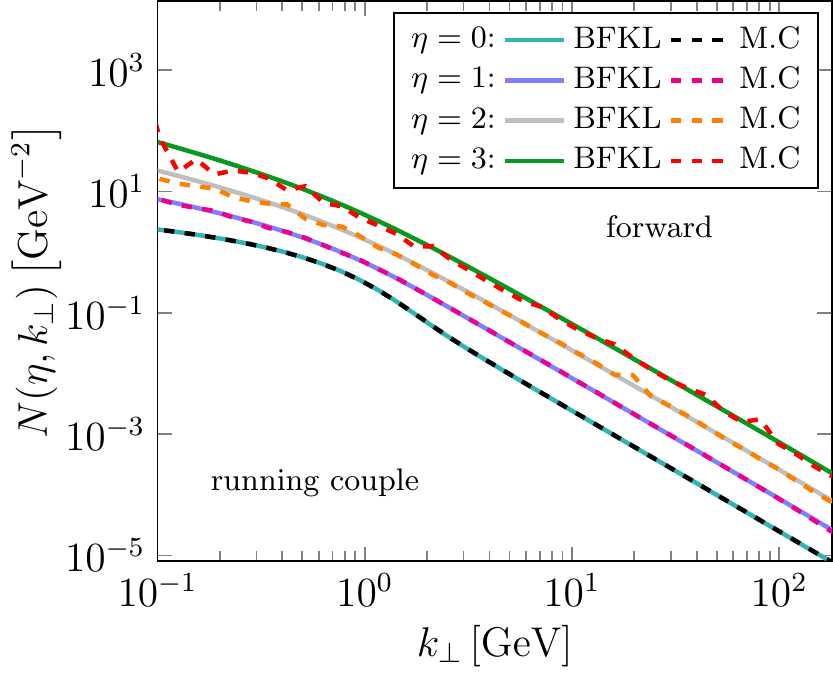}
\caption{Comparison of the gluon $k_\perp$ distributions constructed from the forward evolution approach with the numerical solutions of the BFKL equation at different rapidities. The left and right plots show the results for the standard BFKL evolution in the fixed coupling case and the running coupling cases respectively. We have explicitly checked that those results are independent of the infrared cut-off $\mu$ as long as it is sufficiently small.}
\label{fig:bfklvsmc}
\end{figure}

We repeat the procedure outlined above until $\eta_{i+1}$ reaches the maximum cut-off value $\eta_{\rm max}$. Once the whole cascade is generated, we are ready to reconstruct the gluon $k_\perp$ distribution at arbitrary rapidity, and compare it with the numerical solutions of BFKL equation. For a given $\eta$, we select the event with two adjacent splittings occurring at $\eta_i$ and $\eta_{i+1}$ that satisfies the condition $\eta_i<\eta<\eta_{i+1}$. The event associated with the weight given above is recorded. The gluon distribution $N (\eta, k_\perp)$ is simply the sum of all the re-weighted events. Notice that the final re-weighting factor of this particular event is the product of all those re-weighting factors associated with previous branchings that happen {\it before} $\eta_i$. The re-weighting factor associated with the branching at $\eta_{i}$ is not included. The gluon distributions constructed from the parton cascade are presented in the left panel of Fig.~(\ref{fig:bfklvsmc}), and compared with the numerical solutions of the standard BFKL equation. As one can see, a full agreement between two approaches has been reached. 

The extension to the running coupling case is straightforward in the so-called parent dipole prescription, where the scale is chosen to be 
\begin{equation}
\alpha_s(k_{\perp,i}^2) =\frac{1}{\beta_0 \ln \left[(k_{\perp,i}^2+\mu_0^2)/\Lambda_{\rm QCD}^2\right]},
\end{equation}
with $\beta_0 =(33-2 N_f )/(12\pi) $, $N_f=3$, $\Lambda^2_{\text{QCD}} =0.0578$ GeV$^2$ and $\mu_0^2= 0.942 $ GeV$^2$. Here, $\mu_0$ is introduced to avoid the Landau pole. Therefore, the coupling constant is frozen to $\alpha_s \approx 0.5$ in the infrared region. One first needs to replace the fixed coupling constant $\alpha_s$ in Eq.~(\ref{14}) with $\alpha_s(k_{\perp,i}^2)$. Then, the re-weighting factor is correspondingly modified as,
\begin{equation}
{\cal W}_ {\rm rc} (k_{\perp,i}, k_{\perp,i+1}) =\frac{ \alpha_s (k_{\perp,i+1}^2 )\ln (P_\perp^2/\mu^2) }{\alpha_s (k_{\perp,i}^2 ) \ln (k_{\perp,i}^2/\mu^2) }.
\end{equation}
With these recipes, parton cascade can be generated following the procedure described above. As shown in the right panel of Fig.~\ref{fig:bfklvsmc}, the gluon $k_\perp$ distribution obtained from the Monte Carlo approach matches the numerical results perfectly in the running coupling case as well.

\begin{figure}[htb]
\centering
\includegraphics[width=0.4\textwidth]{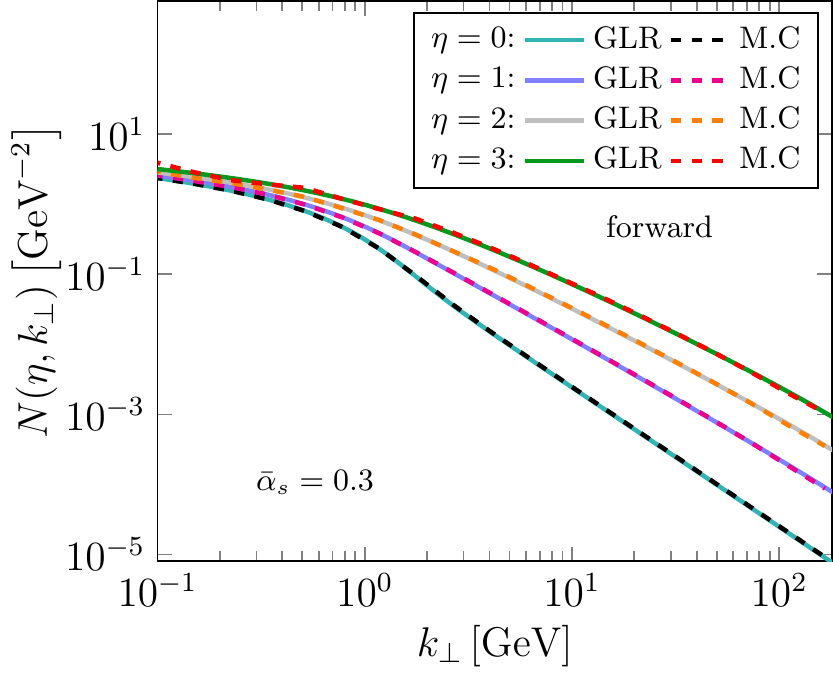}
\includegraphics[width=0.4\textwidth]{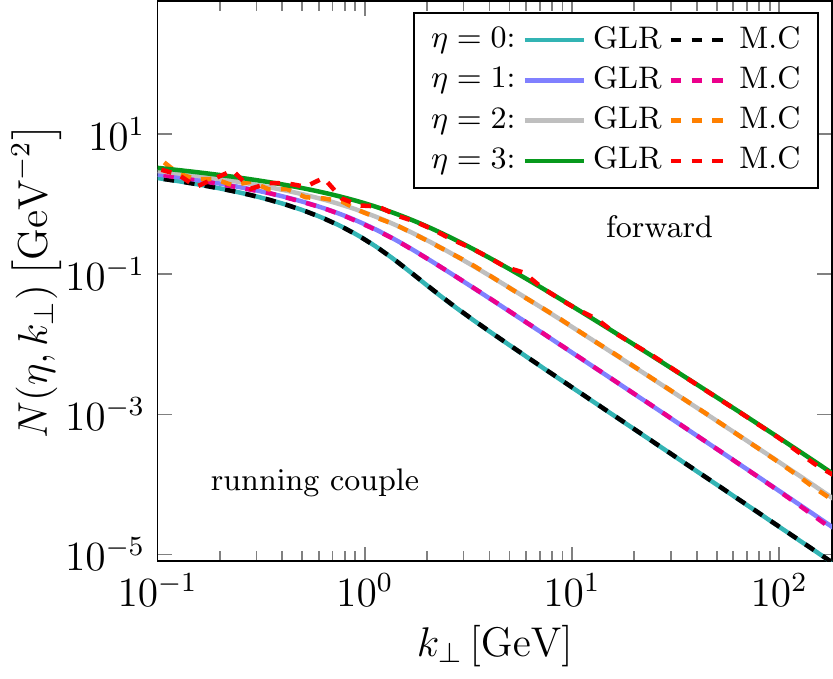}
\caption{Comparison of the gluon $k_\perp$ distributions obtained from the forward evolution approach with the numerical solutions of the GLR equation at different rapidities. The left and right plots show the results for the standard GLR evolution in the fixed coupling case and the running coupling case respectively. }
\label{fig:bkvsmc}
\end{figure}

Now we generalize this algorithm to the saturation case, i.e., the formulation of forward evolution for the GLR equation. First, for a given $\eta_i$ from the previous evolution step or the initial condition, the next $\eta_{i+1}$ can be generated by solving the following equation with the non-Sudakov factor incorporating the saturation term, 
\begin{equation}
\mathcal R = \exp\left[- \bar \alpha_s \int ^{\eta_{i+1}}_{\eta_i} \dd \eta ^\prime \left(\ln \frac{ k_{\perp,i}^2}{\mu ^2} + N(\eta ^\prime, k_{\perp,i}) \right) \right], \label{etai1}
\end{equation}
where the numerical solutions of the BK equation are used as the input for the gluon distribution $N(\eta ^\prime, k_{\perp,i} )$. In the practical simulation, we again employ a veto algorithm to speed up the generation of $\eta_{i+1}$ as described in Appendix B. The re-weighting function also needs to be modified accordingly for the saturation case. It is then given by
\begin{equation}
\mathcal W(\eta_i, \eta_{i+1}; k_{\perp,i})
= \frac{\int_{\eta_{i}}^{\eta_{i+1}} \dd \eta \ln (P_\perp^2/\mu^2)}{\int^{\eta_{i+1}}_{\eta_i} \dd\eta \left[\ln (k_{\perp,i}^2/\mu^2) + N(\eta, k_{\perp,i}) \right]}.
\end{equation}
The rest recipes for the Monte Carlo implementation of both the fixed coupling and the running coupling GLR equation are the same as those for the BFKL equation. 

The gluon $k_\perp$ distributions at different rapidities reconstructed from the parton shower are presented in Fig.~\ref{fig:bkvsmc}, and are compared to the numerical solutions of the GLR equation. A full agreement between two approaches has been reached for both the fixed coupling (left panel) and running coupling (right panel) cases. 

\section{Backward evolution}

The forward evolution procedure developed in the previous section is a direct way of solving the small $x$ evolution equation. However, the forward evolution is rather time-consuming, since the kinematics constructed from the initial state cascade do not have the right values that allow the generation of a hard scattering process most of the time. Many configurations produced by the forward evolution have to be rejected, leading to low efficiency. A more efficient procedure for generating the initial state parton shower is the backward evolution scheme~\cite{Sjostrand:1985xi,Bengtsson:1986gz,Marchesini:1987cf}, which has been utilized in standard Monte Carlo programs. In a backward evolution approach, the hard scattering process is first created with the initial parton momentum distributed according to the parton distribution functions. Then, the initial state cascade is generated by going backward from the hard scattering process towards the beam particles. 

The first step in the formulation of backward evolution is to sample $k_{\perp,i+1}$ at the rapidity $\eta_{i+1}$ that is fixed according to the kinematics of the generated hard scattering process. The value of $k_{\perp,i+1}$ is randomly chosen according to the probability distribution $N(\eta_{i+1},k_{\perp,i+1})$ which has to be determined beforehand by numerically solving the GLR equation. The next step is to generate $\eta_{i}$ using a modified non-Sudakov form factor.

We now derive the non-Sudakov form factor associated with backward evolution for the GLR equation by closely following the DGLAP case (see, for example~\cite{Ellis:1996mzs}). Let us start by defining $\dd F$ as the fraction of gluons at $(\eta_{i+1},k_{\perp,i+1})$ that come from branching between $(\eta_{i+1},\eta_{i})$. Then, the fraction of those that do not branch between $\eta_{i+1}$ and $\eta_{i}$ is,
\begin{equation}
\Pi (\eta_{i+1}, \eta_i; k_{\perp,i+1}) = 1-\int ^{\eta_{i+1}}_{\eta_i} \dd F.
\end{equation}
According to integral form of the folded GLR equation in Eq.~(\ref{eq:inbk}), the number of gluons produced from the branching between $(\eta_{i+1},\eta_{i})$ is given by,
\begin{eqnarray}
N(\eta_{i+1}, k_{\perp,i+1} ) \dd F&=& \dd \eta_i \frac{\Delta(\eta_{i+1}, k_{\perp,i+1})}{\Delta(\eta_i,k_{\perp,i+1})}\frac{\bar \alpha_s }{\pi} \int_\mu \frac{\dd^2 l_{\perp}}{l_\perp^2} N(\eta_i, k_{\perp,i+1}+ l_\perp)
\nonumber \\&=& d\eta_i \frac{\partial}{\partial \eta_i} \left [ \frac{\Delta(\eta_{i+1},k_{\perp,i+1}) N(\eta_i ,k_{\perp,i+1})}{\Delta(\eta_i,k_{\perp,i+1})} \right ],
\label{sd1}
\end{eqnarray}
where we have employed the differential form of the folded GLR equation in Eq.~(\ref{folded}) to get the result in the second line. Performing the integration in the above equation, one obtains,
\begin{eqnarray}
\Pi (\eta_{i+1}, \eta_i; k_{\perp,i+1})=\frac{\Delta(\eta_{i+1},k_{\perp,i+1}) N(\eta_i,k_{\perp,i+1})}{\Delta(\eta_i,k_{\perp,i+1}) N(\eta_{i+1},k_{\perp,i+1})},
\end{eqnarray}
which is the backward evolution form factor describing the probability for no radiation in the rapidity region $[\eta_{i+1}, \eta_i]$. This form factor can be cast into a different form. It is convenient to re-express 
the Eq.~(\ref{eq:inbk}) as,
\begin {eqnarray}
% \frac{\dd \frac{ N(\eta,k_{\perp,i+1})}{\Delta(\eta,k_{\perp,i+1})} }{\frac{ N(\eta,k_{\perp,i+1})}{\Delta(\eta,k_{\perp,i+1})} }
\dd \ln \frac{ N(\eta,k_{\perp,i+1})}{\Delta(\eta,k_{\perp,i+1})}
&=& \dd \eta \frac{\bar \alpha_s}{\pi} \int_\mu \frac{\dd^2 l_{\perp}}{l_\perp^2} \frac{ N(\eta, k_{\perp,i+1}+ l_\perp)}{ N(\eta,k_{\perp,i+1})}.
\end{eqnarray}
% This equation can be integrated between $\eta_i$ and $\eta_{i+1}$ giving,
Carrying out the integration of $\eta$ in the range of $[\eta_i, \eta_{i+1}]$, we obtain,
\begin{eqnarray}
\frac{\Delta(\eta_{i+1},k_{\perp,i+1}) N(\eta_i,k_{\perp,i+1})}{\Delta(\eta_i,k_{\perp,i+1}) N(\eta_{i+1},k_{\perp,i+1})}
&=&\exp \left [-\frac{ \bar \alpha_s}{\pi} \int_{\eta_i}^{\eta_{i+1} } \dd \eta \int_\mu \frac{\dd^2 l_{\perp}}{l_\perp^2} \frac{N(\eta, k_{\perp,i+1}+ l_\perp)}{ N(\eta,k_{\perp,i+1})}\right ],
\label{conection}
\end{eqnarray}
where the new form of the backward evolution form factor is given on the right side of the above equation. Unlike the case for the forward evolution which may be used as a way of solving the GLR equation, the evolved gluon distribution is used as the input to guide the evolution path toward the initial condition at $\eta_0$. The primary aim is to generate the correct distribution of gluons emitted in the initial state cascade. 

Both non-Sudakov forms can be equally well used to generate $\eta_i$ for a given $\eta_{i+1}$ by solving the following equation, 
\begin{equation}
\Pi (\eta_{i+1}, \eta_i; k_{\perp,i+1})= \mathcal R_1.
\end{equation}
The transverse momentum of the radiated gluon sampled in the backward evolution approach is different from that in the forward approach. One should generate $l_{\perp}$ by solving the following equation, 
\begin{equation}
\frac{\bar \alpha_s }{\pi} \int_\mu^{l_\perp} \frac{\dd^2 l'_\perp}{l{'}_\perp^{2}} N(\eta_i, k_{\perp,i+1}+ l'_\perp)= \mathcal R_2 \frac{\bar \alpha_s }{\pi} \int_\mu^{P_\perp} \frac{\dd^2 l'_\perp}{l{'}_\perp^{2}} N(\eta_i, k_{\perp,i+1}+ l'_\perp).
\end{equation}
Notice that, in the backward evolution case, one should not sample $l_\perp$ according to the distribution of $\frac{\bar \alpha_s }{\pi} \int_\mu^{l_\perp} \frac{\dd^2 l'_\perp}{l{'}_\perp^2}$ that is used in the forward evolution approach. Once again, a veto algorithm is employed in our practical implementation to make this sampling procedure more efficient. As mentioned before, due to the mismatch between the phase spaces of real and virtual contributions, the unitarity is violated in the small $x$ evolution. As a consequence, the generated event has to be re-weighted after each branching in the backward evolution method as well. The re-weighting factor associated with backward evolution is the ratio of the fraction of gluons that come from branchings in the region of $[\eta_i,\eta_{i+1}]$ and the fraction of gluons that vanish in the region of $[\eta_i,\eta_{i+1}]$ due to the virtual correction and the fusion process. It reads,
\begin{equation}
\mathcal W_{\rm back}(\eta_{i+1}, \eta_{i}; k_{\perp,i+1}, k_{\perp,i})
= \frac{\int^{\eta_{i+1}}_{\eta_i} \dd\eta \left[\ln (k_{\perp,i}^2/\mu^2) + N(\eta, k_{\perp,i}) \right]}{ \int_{\eta_{i}}^{\eta_{i+1}} \dd \eta \ln (P_\perp^2/\mu^2) }.
\end{equation}
The procedure outlined above is repeated until $\eta_i$ is smaller than $\eta_0$. The last step of the simulation is to construct four momenta of the radiated gluons. It is worth mentioning that the minus component of the $t$-channel gluon's four momentum can only be reconstructed after the full cascade has been generated. By going from the last $t$-channel gluon (closet to the nucleus), which has the vanishing minus component, forward in the cascade to the hard scattering process, the true minus component of the $t$-channel gluons are constructed. It is straightforward to extend to the running coupling case as it has been done in the previous section. The corresponding re-weighting factor and non-Sudakov form factor are given by,
\begin{equation}
\mathcal W_{\rm back, rc}( \eta_{i+1},\eta_i; k_{\perp,i+1}, k_{\perp,i})
=
 \frac{
\int^{\eta_{i+1}}_{\eta_i} \dd\eta
\left[\ln (k_{\perp,i}^2/\mu^2) + N(\eta, k_{\perp,i}) \right] 
}{ \int_{\eta_{i}}^{\eta_{i+1}} \dd \eta \ln (P_\perp^2/\mu^2) }
\frac{ \alpha_s ( k_{\perp,i})}{ \alpha_s ( k_{\perp,i+1})} ,
\end{equation}
and
\begin{eqnarray}
\Pi _{\rm rc} (\eta_{i+1}, \eta_i; k_{\perp,i+1})
&=&\exp \left [-\frac{\bar \alpha_s \left(k_{\perp,i+1}^2 \right )}{\pi} \int_{\eta_i}^{\eta_{i+1} } \dd \eta \int_\mu \frac{\dd^2 l_{\perp}}{l_\perp^2} \frac{N(\eta, k_{\perp,i+1}+ l_\perp)}{ N(\eta,k_{\perp,i+1})} \right ].
\label{conection}
\end{eqnarray}

\begin{figure}[htb]
\centering
\includegraphics[width=0.4\textwidth]{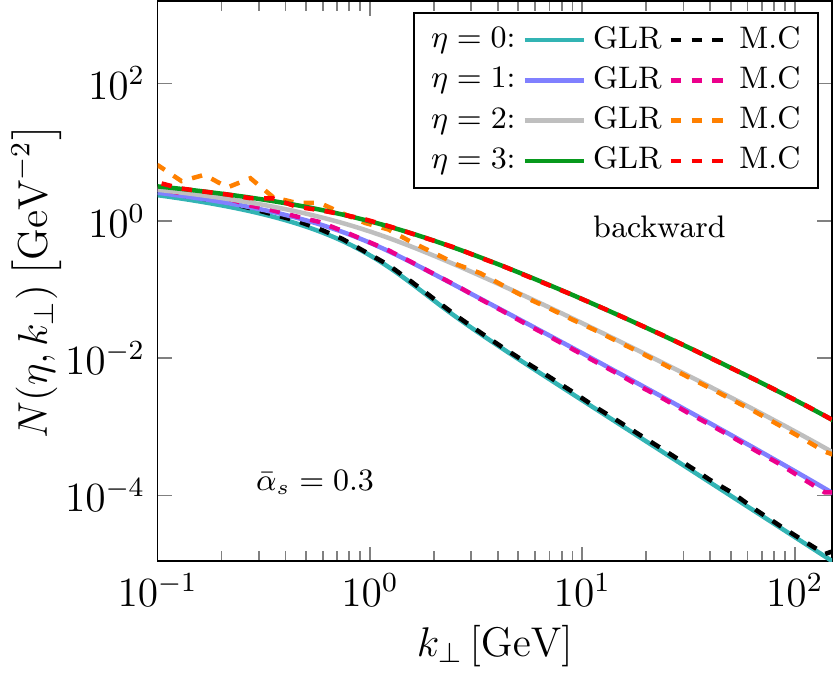}
\includegraphics[width=0.4\textwidth]{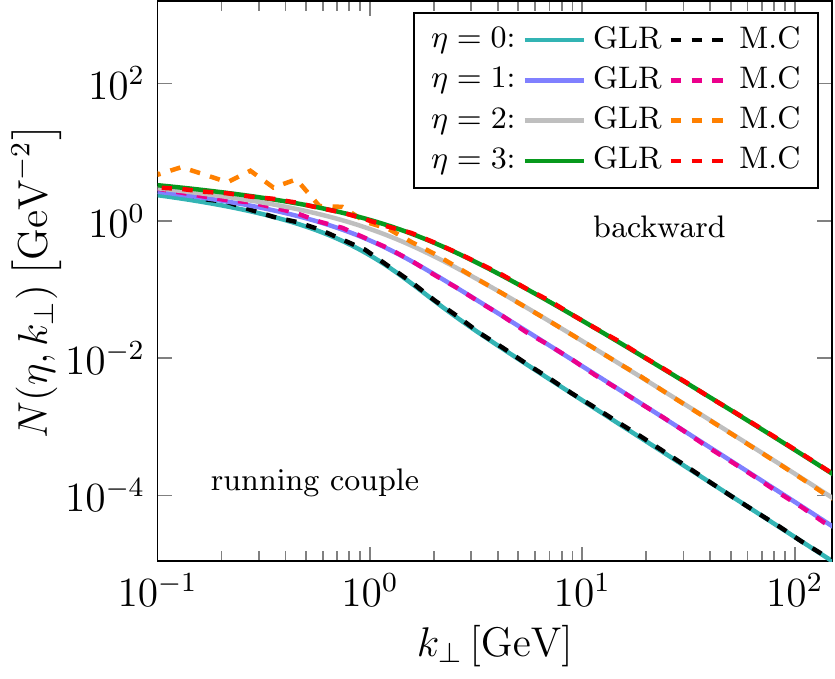}
\caption{Compassion of the gluon $k_\perp$ distributions obtained   from the backward evolution approach with the numerical solutions of the GLR equation at different rapidities. The left and right plots show the results for the standard GLR evolution and the running coupling case respectively.}
\label{fig:backward}
\end{figure}

In Fig.~\ref{fig:backward}, we compare the gluon $k_\perp$ distributions at different rapidities generated from the backward evolution with the numerical solutions of the GLR equation. One can see that the backward approach perfectly reproduces the numerical solutions as shown in in Fig.~\ref{fig:backward}. Therefore, the branching algorithm for backward evolution presented in this section passes the important consistency check. 

\section{Conclusion}

We have developed a Monte Carlo algorithm for simulating the initial state parton branching in the small $x$ region. The underlying parton branching equation employed in our formulation is the GLR equation. To the best of our knowledge, this is the first time that a practical parton shower generator including saturation effect has been constructed. We first derived a folded form of the GLR equation and the associated non-Sudakov form factor which is the starting point for the Monte Carlo implementation. With the derived non-Sudakov form factor, a forward evolution scheme which can be viewed as a direct way of solving the evolution equation, is developed and is shown to reproduce the numerical solutions of the GLR equation. As a more efficient procedure, the backward evolution approach is also presented. It yields the same results as the forward approach produces as expected. 

To build a full hadron-level Monte Carlo generator for simulating events in $eA$ collisions, the next step is to perform the hadronization using multi-purpose generators such as PYTHIA~\cite{Sjostrand:2006za} after parton-level events have been generated. Such an event generator can be used for the description of fully exclusive observables in $eA$ collisions or the forward region of $pA$ collisions, as well as for EIC impact studies. We leave this for future works. Apart from the studies of exclusive events, another advantage of the Monte Carlo method over the conventional analytical approach is that four-momentum conservation is explicitly imposed in each step of the parton branching. As pointed out in Refs.~\cite{Xiao:2018zxf, Kang:2019ysm,Liu:2020mpy,Liu:2022xsc,Shi:2021hwx,Liu:2022ijp}, it is crucial to take into account the exact kinematics effect to correctly describe particle production near the threshold region. As long as the momentum conservation is kept, the threshold resummation is automatically carried out in the parton branching algorithm, whereas it is quite a nontrivial task to achieve in the analytical calculations. 

There still remains much room for further theoretical progress in the development of a small $x$ parton shower generator. For example, it is important to investigate whether a Monte Carlo implementation of the BK equation is possible or not. To this end, one has to go beyond the triple pomeron vertex approximation represented by the non-linear term in the GLR equation. The multiple re-scattering between the emitted gluons and the dense medium inside the large nucleus has to be properly treated in the parton shower generator. On the other hand, the linear polarization of small $x$ gluons~\cite{Metz:2011wb} needs to be taken into account in future updates as well. Moreover, for the case of the processes involving multiple well-separated hard scales, a joint small $x$ and $k_t$ resummation~\cite{Zhou:2016tfe,Xiao:2017yya,Zhou:2018lfq,Taels:2022tza,Caucal:2022ulg,Goda:2022wsc} needs to be performed. The extension of the parton branching algorithm to embody $k_t$ resummation is crucial for the phenomenological studies of back-to-back two particles/jets production processes in $eA$ collisions. We will address these issues in a future publication. 

{\it Acknowledgments:}
We thank Hai-tao Li and Shan-shan Cao for helpful discussions.
This work has been supported by the National Natural Science Foundation of China under Grant No. 1217511.
S.Y.W. is also supported by the Taishan fellowship of Shandong Province for junior scientists. 
%%%%%%%%%%%%%%%%%%%%%%%%%%%%%%%%%%%%%%%%
%%%%%
%%%%%%%%%%%%%%%%%%%%%%%%%%%%%%%

% \section*{Appendix}

% \global\long\def\theequation{A\arabic{equation}}%
%  \setcounter{equation}{0} \label{A} 

\appendix
%%dummy comment inserted by tex2lyx to ensure that this paragraph is not empty%dummy comment inserted by tex2lyx to ensure that this paragraph is not empty
%

\section{The different forms of the virtual part of the BFKL kernel}
It is easy to show the equivalence of the following two integrals using the dimensional regularization method.
 \begin{eqnarray}
\int\frac{\dd^{2}\lperp}{(\kperp-\lperp)^{2}} \frac{\kperp^2}{2\lperp^{2}}
=
\int^{|\kperp|} \frac{\dd^{2}\lperp}{\lperp^{2}}.\label{a1}
\end{eqnarray}
With the help of the standard Feynman parameters approach, we obtain,
\begin{eqnarray}
 &  &  \!\!\!\!\!\!\!\!\!
 (\frac{\mu^{2}e^{\gamma_{E}}}{4\pi})^{\epsilon}\int\frac{\dd^{2-2\epsilon}\lperp}{(2\pi)^{2-2\epsilon}}\frac{\kperp^{2}}{\lperp^{2}(\kperp-\lperp)^{2}}= (\frac{\mu^{2}e^{\gamma_{E}}}{4\pi})^{\epsilon}\int_{0}^{1}\dd x \int \frac{\dd^{2-2\epsilon}\lperp}{(2\pi)^{2-2\epsilon}}\frac{\kperp^{2}}{\left[(1-x)\lperp^{2}+x(\kperp-\lperp)^{2}\right]^{2}}\nonumber \\
 & = & (\frac{\mu^{2}e^{\gamma_{E}}}{4\pi})^{\epsilon}\int_{0}^{1}\dd x\int\frac{\dd^{2-2\epsilon}\lperp}{(2\pi)^{2-2\epsilon}}\frac{\kperp^{2}}{\left[\lperp^{2}+(1-x)x\kperp^{2}\right]^{2}}= \frac{2}{4\pi } (-\frac{1}{\epsilon} +\ln \frac{\kperp ^2}{\mu ^2} ) +\mathcal{O}(\epsilon).
\end{eqnarray}
On the other hand, one has,
\begin{eqnarray}
 (\frac{\mu^{2}e^{\gamma_{E}}}{4\pi})^{\epsilon}\int\frac{\dd^{2-2\epsilon}\lperp}{(2\pi)^{2-2\epsilon}}\frac{1}{\lperp^{2}} \theta (|\kperp| -|l_\perp|) =  (\frac{\mu^{2}e^{\gamma_{E}}}{4\pi})^{\epsilon} 
 \frac{1}{(2\pi)^{2-2\epsilon}}
 \frac{2\pi ^{1-\epsilon}}{\Gamma(1-\epsilon)} \frac{1}{-2\epsilon} \frac{1}{\kperp^{2\epsilon}}
 = \frac{1}{4\pi}(-\frac{1}{\epsilon} +\ln \frac{\kperp ^2}{\mu ^2} )  +\mathcal{O}(\epsilon).
\end{eqnarray}
Therefore,  the relation given in Eq.~(\ref{a1}) holds. We also checked this relation using the  different regularization prescriptions and confirmed the equivalence.

\section{The veto algorithm in the Monte-Carlo simulation}

In this appendix, we discuss the veto algorithm in more details. Let us first recall how to sample a distribution  $f(x)$ in the Monte-Carlo simulation. First, we calculate the integral $F(x) = \int^{x}_{x_{\min}} dx' f(x')$ and the normalization factor $C=\int_{x_{\min}}^{x_{\max}} dx' f(x')$ where $x_{\min}$ and $x_{\max}$ define the $x$ regime where the sampled events reside. We can simply generate a random number ${\cal R} \in [0,1]$ and obtain the sampled event of $x$ by solving $x=F^{-1} (C {\cal R})$ where $F^{-1}$ is the inverse function of $F$. We replicate the same procedure to generate more events. Statistically, the sampled events automatically satisfy the $f(x)$ distribution. 

However, if the integral  $f(x)$ can not be carried out analytically, the above approach does not apply. The  solution to this problem is the veto algorithm. The essential point is to find a simple analytically integrable function, which is always larger than the desired distribution  $f(x)$.
As an example, we explain how to generate the value of $k_\perp$ at the initial rapidity with a veto algorithm.  The gluon $k_\perp$ distribution  ${ N} (\eta_0=0, k_\perp)$ is computed in the MV model.  First, we construct a test function $f(k_\perp) = {\cal C}/(k_\perp^2 + Q_0^2)$. By properly choosing the ${\cal C}$ and $Q_0$ parameters, we make sure that $f(k_\perp) \ge { N} (0, k_\perp)$ in the whole $k_\perp$ region of interest. Next, we generate a value of  $k_\perp$ according to the test function.  Third, this event is accepted with the probability according to the ratio  $N (0, k_\perp)$ and $f(k_\perp)$.  Otherwise, the event is rejected.

% \begin{align}
% \frac{1}{N_{\rm events}} \frac{dN_{\rm events}}{d^2 k_\perp} = \frac{1}{\int d^2k_\perp {\cal N} (x_0, k_\perp)} {\cal N} (x_0, k_\perp),
% \end{align}

The non-Sudakov form factor associated with the GLR equation involves  the gluon distribution function which is not an analytically integrable function. It is necessary to invoke a veto algorithm~\cite{Lonnblad:2012hz} for selecting the value of $\eta_{i+1}$ as well. The gluon $k_\perp$ distribution ${ N} (\eta, k_\perp)$ can be replaced with the test function $f(k_\perp)$ that satisfies $f(k_\perp)>{ N} (\eta, k_\perp)$ in the entire kinematic region and the entire rapidity region of interest. Thus a simple analytically calculable form of the non-Sudakov form factor is obtained. The algorithm is described as the follows.
\begin{itemize}
\item  We generate a $\eta_{i+1}$ by solving the following equation, 
\begin{equation}
\mathcal R_1
= \exp\left[- \bar \alpha_s (\eta_{i+1} - \eta_i) \left(  \ln \frac{ k_{\perp,i} ^2}{\mu ^2}  +  f(k_{\perp,i})  \right) \right].
\end{equation}
\item This generated event is accepted with the probability of ${\cal P} = \left[ \ln \frac{ k_{\perp,i} ^2}{\mu ^2}  +  N(\eta_{i+1}, k_{\perp,i}) \right] \Big/ \left[\ln \frac{ k_{\perp,i} ^2}{\mu ^2}  +  f(k_{\perp,i}) \right]$. 
% Accepting the generated $\eta_{i+1}$ with the probability according to the ratio . 
\item If the generated event is rejected, we first replace $\eta_{i}$ with $\eta_{i+1}$ generated from the first step and then go back to the first step to re-generate a new $\eta_{i+1}$. We repeat this procedure until a $\eta_{i+1}$ is accepted.
\item After we finally obtained an accepted $\eta_{i+1}$, we can then proceed to generate the transverse momentum of the radiated gluon as described in Sec.~III.
\end{itemize}

\bibliographystyle{apsrev4-1}
\bibliography{ref.bib}

\end{document}